\documentstyle[pra,aps,preprint]{revtex}
\tightenlines

\begin{document}
\draft
\title{Periodic Orbit Quantization: How to Make Semiclassical Trace 
Formulae Convergent\footnote{Contribution to 
``Festschrift in honor of Martin Gutzwiller'', eds.\ A.\ Inomata et al., 
to be published in {\em Foundations of Physics}.}}
\author{J\"org Main and G\"unter Wunner}
\address{Institut f\"ur Theoretische Physik I, 
Universit\"at Stuttgart, D-70550 Stuttgart, Germany}
\date{\today}
\maketitle

\begin{abstract}
Periodic orbit quantization requires an analytic continuation of 
non-con\-ver\-gent semiclassical trace formulae.
We propose two different methods for semiclassical quantization.
The first method is based upon the harmonic inversion of semiclassical 
recurrence functions.
A band-limited periodic orbit signal is obtained by analytical frequency 
windowing of the periodic orbit sum.
The frequencies of the periodic orbit signal are the semiclassical 
eigenvalues, and are determined by either linear predictor, 
Pad\'e approximant, or signal diagonalization.
The second method is based upon the direct application of the 
Pad\'e approximant to the periodic orbit sum.
The Pad\'e approximant allows the resummation of the, typically
exponentially, divergent periodic orbit terms.
Both techniques do not depend on the existence of a symbolic dynamics,
and can be applied to bound as well as to open systems.
Numerical results are presented for two different systems with chaotic
and regular classical dynamics, viz.\ the three-disk scattering system 
and the circle billiard.
\end{abstract}

\pacs{PACS numbers: 05.45.$-$a, 03.65.Sq}

\section{Introduction}
\label{intro}
Semiclassical periodic orbit quantization is a nontrivial problem for 
the reason that Gutzwiller's trace formula \cite{Gut67,Gut90} for chaotic
systems and the Berry-Tabor formula \cite{Ber76} for integrable systems
do not usually converge in those regions where the physical eigenenergies 
or resonances are located.
Various techniques have been developed to circumvent the convergence 
problem of periodic orbit theory.
Examples are the cycle expansion technique \cite{Cvi89}, the Riemann-Siegel 
type formula and pseudo-orbit expansions \cite{Ber90}, surface of section
techniques \cite{Bog92}, and a quantization rule based on a semiclassical
approximation to the spectral staircase \cite{Aur92}.
These specific techniques have proven to be very efficient for individual
systems with special properties, e.g., the cycle expansion for hyperbolic 
systems with an existing symbolic dynamics.
The other methods mentioned have been used for the calculation of bound 
spectra of specific systems.

Recently, an alternative method based upon filter-diagonalization (FD) 
has been introduced for the analytic continuation of the semiclassical 
trace formula \cite{Mai97a,Mai98a}.
The FD method requires knowledge of the periodic orbits up to a given maximum 
period (classical action), which depends on the mean density of states.
The semiclassical eigenenergies or resonances are obtained by 
{\em harmonic inversion\/} of the periodic orbit recurrence signal.
The FD method can be generally applied to both open and bound systems 
and has also proven to be a powerful tool, e.g., for the calculation of 
semiclassical transition matrix elements \cite{Mai99a} and the quantization 
of systems with mixed regular-chaotic phase space \cite{Mai99b}.
For a review on periodic orbit quantization by harmonic inversion see 
\cite{Mai99c}.

In this paper we present two different techniques to make semiclassical
periodic orbit sums convergent.
The first method is an advanced version of harmonic inversion adapted
to the special structure of periodic orbit signals given as sums of
$\delta$ functions \cite{Mai00}.
The semiclassical signal, in action or time, corresponds to a ``spectrum''
or response in the frequency domain that is composed of a huge, in principle
infinite, number of frequencies.
To extract these frequencies and their corresponding amplitudes is a 
nontrivial task.
In previous work \cite{Mai97a,Mai98a,Mai99c} the periodic orbit signal
has been harmonically inverted by means of FD \cite{Wal95,Man97a,Man97b}
which is designed for the analysis of time signals given on an equidistant 
grid.
The periodic orbit recurrence signal is represented as a sum over usually 
unevenly spaced $\delta$ functions. 
A smooth signal, from which evenly spaced values can be read off, is obtained 
by a convolution of this sum with, e.g., a narrow Gaussian function.
The disadvantages of this approach are twofold.
Firstly, FD acts on this signal more or less like a ``black box''
and, as such, does not lend itself to a detailed understanding of 
semiclassical periodic orbit quantization.
Secondly, the smoothed semiclassical signal usually consists of a huge 
number of data points.
The handling of such large data sets, together with the smoothing, may 
lead to significant numerical errors in results for the semiclassical 
eigenenergies and resonances.
Here, we propose alternative techniques for the harmonic inversion of
the periodic orbit recurrence signal that avoid these problems.
In a first step we create a shortened signal which is constructed 
from the original signal and designed to be correct only in a window, 
i.e., a short frequency range of the total band width.
Because the original signal is given as a periodic orbit sum of 
$\delta$ functions, this ``filtering'' can be performed analytically 
resulting in a band-limited periodic orbit signal with a relatively small 
number of equidistant grid points.
In a second step the frequencies and amplitudes of the band-limited signal
are determined from a set of nonlinear equations.
To solve the nonlinear system, we introduce three different processing 
methods, viz.\ linear predictor (LP), Pad\'e approximant (PA), and 
signal diagonalization (SD).
It is important to note that these processing methods would not have yielded
numerically stable solutions if the signal had not first been band-limited 
by the windowing (filtering) procedure.
Furthermore, this separation of the harmonic inversion procedure into 
various steps may elucidate a clearer picture of the periodic orbit 
quantization method itself, and even provides more accurate results than 
previous calculations \cite{Mai98a,Mai99c} using FD.

The second method is the direct application of the Pad\'e approximant to 
slowly convergent and/or divergent periodic orbit sums \cite{Mai99d}.
In the former or the latter case, the PA either significantly increases 
the convergence rate, or analytically continues the {\it exponentially}
divergent series, respectively.
The PA is especially robust for resumming diverging series in many 
applications in mathematics and theoretical physics \cite{Bak75}.
An important example is the summation of the divergent Rayleigh-Schr\"odinger
quantum mechanical perturbation series, e.g., for atoms in electric 
\cite{CiV82} and magnetic \cite{Bel89} fields.
In periodic orbit theory the PA has been applied to cycle-expanded Euler 
products and dynamical zeta functions \cite{Eck93}.
It should be noted that the Pad\'e approximant is applied in both methods
in a completely different context, namely, in the first method as a tool for
signal processing \cite{Mai00,Bel00}, and in the second for the direct 
summation of the periodic orbit terms in the semiclassical trace formulae.

In Sec.\ \ref{method1:sec} we present our first method to make semiclassical
trace formulae convergent.
After briefly reviewing the general idea of periodic orbit quantization 
by harmonic inversion in Sec.\ \ref{po_quant:sec} we construct, in 
Sec.\ \ref{analyt_dec:sec}, the band-limited periodic orbit signal which 
is analyzed, in Sec.\ \ref{hi:sec}, with the help of either LP, PA, or SD.
In Sec.\ \ref{method2:sec} we introduce our second method for semiclassical
quantization, viz.\ the direct application of the Pad\'e approximant to 
the periodic orbit sum.
In Sec.\ \ref{results:sec} we present and compare results for the three-disk 
repeller and the circle billiard as physical examples of relevance.
A few concluding remarks are given in Sec.\ \ref{concl:sec}.

\section{Harmonic inversion of periodic orbit signals}
\label{method1:sec}
\subsection{General remarks}
\label{po_quant:sec}
In order to understand what follows, a brief recapitulation of the basic 
ideas of periodic orbit quantization by harmonic inversion may be useful.
For further details see \cite{Mai99c}.

Following Gutzwiller \cite{Gut67,Gut90} one can write the semiclassical 
response function for chaotic systems in the form
\begin{equation}
 g^{\rm sc}(E) = g^{\rm sc}_0(E)
   + \sum_{\rm po} {\cal A}_{\rm po} {\rm e}^{{\rm i}S_{\rm po}} \; ,
\label{gE_sc}
\end{equation}
where $g^{\rm sc}_0(E)$ is a smooth function and $S_{\rm po}$ 
and ${\cal A}_{\rm po}$ are the classical actions and weights 
(including phase information given by the Maslov index) of the 
periodic orbit (po) contributions.
Equation~(\ref{gE_sc}) is also valid for integrable systems when the periodic
orbit quantities are calculated not with Gutzwiller's trace formula, but with
the Berry-Tabor formula \cite{Ber76} for periodic orbits on rational tori.
The eigenenergies and resonances are the poles of the response function.
Unfortunately, the semiclassical approximation (\ref{gE_sc}) does not 
converge in the region of the poles, and hence one is faced with the problem 
of the analytic continuation of $g^{\rm sc}(E)$ to this region.

As in previous work \cite{Mai97a,Mai98a,Mai99c}, we will make the (weak) 
assumption that the classical system has a scaling property, i.e., the shape 
of periodic orbits is assumed not to depend on a scaling parameter, $w$, 
and the classical action scales as 
\begin{equation}
 S_{\rm po} = ws_{\rm po} \; .
\label{S_po}
\end{equation}
In scaling systems, the fluctuating part of the semiclassical 
response function,
\begin{equation}
   g^{\rm sc}(w)
 = \sum_{\rm po} {\cal A}_{\rm po} {\rm e}^{{\rm i}ws_{\rm po}} \; ,
\label{g_sc}
\end{equation}
can be Fourier transformed readily to yield the semiclassical trace of 
the propagator
\begin{equation}
   C^{\rm sc}(s) 
 = {1 \over 2\pi} \int_{-\infty}^{+\infty} g^{\rm sc}(w)
   {\rm e}^{-{\rm i}sw} {\rm d}w
 = \sum_{\rm po} {\cal A}_{\rm po} \delta\left(s-s_{\rm po}\right) \; .
\label{C_sc}
\end{equation}
The signal $C^{\rm sc}(s)$ has $\delta$ spikes at the positions of the 
classical periods (scaled actions) $s=s_{\rm po}$ of periodic orbits and 
with peak heights (recurrence strengths) ${\cal A}_{\rm po}$, i.e., 
$C^{\rm sc}(s)$ is Gutzwiller's periodic orbit recurrence function.
Consider now the quantum mechanical counterparts of $g^{\rm sc}(w)$ and
$C^{\rm sc}(s)$ taken as the sums over the poles $w_k$ of the Green's 
function,
\begin{equation}
 g^{\rm qm}(w) = \sum_k {d_k \over w-w_k+{\rm i}\epsilon} \; ,
\label{g_qm}
\end{equation}
\begin{equation}
   C^{\rm qm}(s)
 = {1\over 2\pi} \int_{-\infty}^{+\infty} g^{\rm qm}(w)
   {\rm e}^{-{\rm i}sw} {\rm d}w
 = -{\rm i}\sum_k d_k {\rm e}^{-{\rm i}w_k s} \; ,
\label{C_qm}
\end{equation}
with $d_k$ being the residues associated with the eigenvalues.
In the case under study, i.e., density of states spectra, the $d_k$ are the
multiplicities of eigenvalues and are equal to 1 for non-degenerate states.
Semiclassical eigenenergies $w_k$ and residues $d_k$ can now, in principle, 
be obtained by adjusting the semiclassical signal, Eq.~(\ref{C_sc}), to 
the functional form of the quantum signal, Eq.~(\ref{C_qm}), with the 
$\{w_k,d_k\}$ being free, in general complex, frequencies and amplitudes.
This scheme is known as ``harmonic inversion''.
The numerical procedure of harmonic inversion is a nontrivial task, 
especially if the number of frequencies in the signal is large 
(e.g., more than a thousand), or even infinite as is usually the 
case for periodic orbit quantization.
Note that the conventional way to perform the spectral analysis, i.e.,
the Fourier transform of Eq.~(\ref{C_sc}) will bring us back to analyzing
the non-convergent response function $g^{\rm sc}(w)$ in Eq.~(\ref{g_sc}).
The periodic orbit signal (\ref{C_sc}) can be harmonically inverted by
application of FD \cite{Wal95,Man97a,Man97b},
which allows one to calculate a finite and relatively small set of
frequencies and amplitudes in a given frequency window.
The usual implementation of FD requires knowledge of the signal on an 
equidistant grid.
The signal (\ref{C_sc}) is not a continuous function.
However, a smooth signal can be obtained by a convolution of $C^{\rm sc}(s)$ 
with, e.g., a Gaussian function,
\begin{equation}
   C_\sigma^{\rm sc}(s) 
 = {1\over \sqrt{2\pi}\sigma}
   \sum_{\rm po} {\cal A}_{\rm po} {\rm e}^{-(s-s_{\rm po})^2/2\sigma^2} \; .
\label{C_sc_sigma}
\end{equation}
As can easily be seen, the convolution results in a damping of the 
amplitudes, $d_k\to d_k^{(\sigma)}=d_k\exp{(-w_k^2\sigma^2/2)}$.
The width $\sigma$ of the Gaussian function should be chosen sufficiently 
small to avoid an overly strong damping of amplitudes.
To properly sample each Gaussian a dense grid with steps
$\Delta s\approx\sigma/3$ is required.
Therefore, the signal (\ref{C_sc_sigma}) analyzed by FD
usually consists of a large number of data points.
The numerical treatment of this large data set may suffer from rounding
errors and loss of accuracy.
Additionally, the ``black box'' type procedure of harmonic inversion by FD, 
which intertwines windowing and processing, does not provide any opportunity 
to gain a deeper understanding of semiclassical periodic orbit quantization.
It is therefore desirable to separate the harmonic inversion procedure into
two sequential steps:
Firstly, the filtering procedure that does not require smoothing
and, secondly, a procedure for extracting  the frequencies and amplitudes.
In Sec.\ \ref{analyt_dec:sec} we will construct, by analytic filtering, a
band-limited signal which consists of a relatively small number of frequencies.
In Sec.\ \ref{hi:sec} we will present methods to extract the frequencies and 
amplitudes of such band-limited signals.

\subsection{Construction of band-limited signals by analytical filtering}
\label{analyt_dec:sec}
In general, a frequency filter can be applied to a given signal by
application of the Fourier transform \cite{Bel00,Zol91,Bel99}.
The signal is transformed to the frequency domain, e.g., by application
of the fast Fourier transform (FFT) method.
The transformed signal is multiplied with a frequency filter function
$f(w)$ localized around a central frequency, $w_0$.
The frequency filter $f(w)$ can be rather general, typical examples are 
a rectangular window or a Gaussian function.
The filtered signal is then shifted by $-w_0$ and transformed back to the
time domain by a second application of FFT.
The filtered signal consists of a significantly reduced set of frequencies,
and therefore a reduced set of grid points is sufficient for the analysis
of the signal.
This technique is known as ``beam spacing'' \cite{Zol91} or ``decimation'' 
\cite{Bel00,Bel99} of signals.

The special form of the periodic orbit signal (\ref{C_sc}) as a sum of
$\delta$ functions allows for an even simpler procedure, viz.\
analytical filtering.
In the following we will apply a rectangular filter, i.e., $f(w)=1$ 
for frequencies $w \in [w_0-\Delta w,w_0+\Delta w]$, and $f(w)=0$ outside 
the window.
The generalization to other types of frequency filters is straightforward.
Starting from the semiclassical response function (spectrum)
$g^{\rm sc}(w)$ in Eq.~(\ref{g_sc}), which is itself a Fourier transform 
of the ``signal'' (\ref{C_sc}), and using a rectangular window we obtain,
after evaluating the ``second'' Fourier transform, the band-limited (bl) 
periodic orbit signal,
\begin{eqnarray}
     C^{\rm sc}_{\rm bl}(s)
 &=& {1\over 2\pi} \int_{w_0-\Delta w}^{w_0+\Delta w}
     g^{\rm sc}(w) {\rm e}^{-{\rm i}s(w-w_0)} {\rm d}w \nonumber \\
 &=& {1\over 2\pi} \sum_{\rm po} {\cal A}_{\rm po}
     \int_{w_0-\Delta w}^{w_0+\Delta w}
     {\rm e}^{{\rm i}sw_0-{\rm i}(s-s_{\rm po})w} {\rm d}w \nonumber \\
 &=& \sum_{\rm po} {\cal A}_{\rm po} {\sin{[(s-s_{\rm po})\Delta w]}\over
     \pi(s-s_{\rm po})} {\rm e}^{{\rm i}s_{\rm po}w_0} \; .
\label{C_sc_bl}
\end{eqnarray}
The introduction of $w_0$ into the arguments of the exponential functions 
in (\ref{C_sc_bl}) causes a shift of frequencies by $-w_0$ in the frequency 
domain.
Note that $C^{\rm sc}_{\rm bl}(s)$ is a smooth function and can be easily
evaluated on an arbitrary grid of points $s_n<s_{\rm max}$ provided the
periodic orbit data are known for the set of orbits with classical action
$s_{\rm po}<s_{\rm max}$.

Applying now the same filter as used for the semiclassical periodic orbit 
signal to the quantum one, we obtain the band-limited quantum signal
\begin{eqnarray}
     C^{\rm qm}_{\rm bl}(s)
 &=& {1\over 2\pi} \int_{w_0-\Delta w}^{w_0+\Delta w}
     g^{\rm qm}(w) {\rm e}^{-{\rm i}s(w-w_0)} {\rm d}w  \nonumber \\
 &=& -{\rm i} \sum_{k=1}^K d_k {\rm e}^{-{\rm i}(w_k-w_0)s} \; ,
       \quad |w_k-w_0| < \Delta w \; .
\label{C_qm_bl}
\end{eqnarray}
In contrast to the signal $C^{\rm qm}(s)$ in Eq.~(\ref{C_qm}), the band-limited
quantum signal consists of a {\em finite} number of frequencies $w_k$, 
$k=1,\dots,K$, where in practical applications $K$ can be of the order of 
$\sim$~(50-200) for an appropriately chosen frequency window, $\Delta w$.
The problem of adjusting the band-limited semiclassical signal in
Eq.~(\ref{C_sc_bl}) to its quantum mechanical analogue in Eq.~(\ref{C_qm_bl})
can now be written as a set of $2K$ nonlinear equations
\begin{equation}
   C^{\rm sc}_{\rm bl}(n\tau) \equiv c_n
 = -{\rm i} \sum_{k=1}^K d_k {\rm e}^{-{\rm i}w'_kn\tau} \; ,
       \quad n=0,1,\dots,2K-1 \; ,
\label{C_bld}
\end{equation}
for the $2K$ unknown variables, viz.\ the shifted frequencies, 
$w'_k\equiv w_k-w_0$, and amplitudes, $d_k$.
The band-limited signal now becomes  ``short''as it can be evaluated 
on an equidistant grid, $s=n\tau$, with relatively large step width 
$\tau\equiv\pi/\Delta w$.
It is important to note that the number of signal points $c_n$ in 
Eq.~(\ref{C_bld}) is usually much smaller than a reasonable discretization
of the signal $C_\sigma^{\rm sc}(s)$ in Eq.~(\ref{C_sc_sigma}), which is
the starting point for harmonic inversion by FD.
Therefore, the discrete signal points $c_n\equiv C^{\rm sc}_{\rm bl}(n\tau)$ 
are called the ``band-limited'' periodic orbit signal.
Methods to solve the nonlinear system, Eq.~(\ref{C_bld}), are discussed in
Sec.\ \ref{hi:sec} below.

It should also be noted that the analytical filtering in Eq.~(\ref{C_sc_bl})
is not restricted to periodic orbit signals, but can be applied,
in general, to any signal given as a sum of $\delta$ functions.
An example is the high resolution analysis of quantum spectra 
\cite{Mai99c,Mai97b,Gre00}, where the density of states is
$\varrho(E)=\sum_n\delta(E-E_n)$.

\subsection{Harmonic inversion of band-limited signals}
\label{hi:sec}
In this section we wish to solve the nonlinear set of equations
\begin{equation}
 c_n = \sum_{k=1}^K d_k z_k^n \; , \quad n=0,1,\dots,2K-1 \; ,
\label{c_n:eq}
\end{equation}
where $z_k\equiv\exp{(-{\rm i}w'_k\tau)}$ and $d_k$ are, generally complex,
variational parameters.
For notational simplicity we have absorbed the factor of $-{\rm i}$ on the 
right-hand side of Eq.~(\ref{C_bld}) into the $d_k$'s with the understanding 
that this should be corrected for at the end of the calculation.
We assume that the number of frequencies in the signal is relatively small
($K\sim 50$ to $200$).
Although the system of nonlinear equations is, in general, still 
ill-conditioned, frequency filtering reduces the number of signal points, 
and hence the number of equations.
Several numerical techniques, that otherwise would be numerically unstable,
can now be applied successfully.
In the following we employ three different methods, viz.\ linear 
predictor (LP), Pad\'e approximant (PA), and signal diagonalization (SD).

\subsubsection{Linear Predictor}
The problem of solving Eq.~(\ref{c_n:eq}) has already been addressed in the 
18th century by Baron de Prony \cite{Prony}, who converted the nonlinear set
of equations (\ref{c_n:eq}) to a linear algebra problem.
Today this method is known as linear predictor (LP).
Our method strictly applies the procedure of LP except with one essential 
difference; the original signal $C^{\rm sc}(s)$ is replaced with its 
band-limited counterpart $c_n \equiv C^{\rm sc}_{\rm bl}(n\tau)$.

Equation~(\ref{c_n:eq}) can be written in matrix form for the signal points
$c_{n+1}$ to $c_{n+K}$,
\begin{equation}
   \left(\begin{array}{c} c_{n+1}\\ \vdots \\ c_{n+K} \end{array} \right)
 = \left(\begin{array}{ccc}
      z_1^{n+1} & \cdots & z_K^{n+1} \\
      \vdots & \ddots & \vdots \\
      z_1^{n+K} & \cdots & z_K^{n+K}
   \end{array} \right)
   \left(\begin{array}{c} d_1 \\ \vdots \\ d_K \end{array} \right) \; .
\label{c_n_matrix}
\end{equation}
From the matrix representation (\ref{c_n_matrix}) it follows that
\begin{equation}
c_n = \left( z_1^n \cdots z_K^n \right)
      \left(\begin{array}{ccc}
          z_1^{n+1} & \cdots & z_K^{n+1} \\
          \vdots & \ddots & \vdots \\
          z_1^{n+K} & \cdots & z_K^{n+K}
      \end{array} \right)^{-1}
      \left(\begin{array}{c} c_{n+1} \\ \vdots \\ c_{n+K} \end{array} \right)
    = \sum_{k=1}^K a_k c_{n+k} \; ,
\label{a_k_def}
\end{equation}
which means that every signal point $c_n$ can be ``predicted'' by a linear
combination of the $K$ subsequent points with a fixed set of coefficients 
$a_k$, $k=1,\dots,K$.
The first step of the LP method is to calculate these coefficients.
Writing Eq.~(\ref{a_k_def}) in matrix form with $n=0,\dots,K-1$, we obtain 
the coefficients $a_k$ as the solution of the linear set of equations,
\begin{equation}
   \left(\begin{array}{ccc}
       c_{1} & \cdots & c_{K} \\
       \vdots & \ddots & \vdots \\
       c_{K} & \cdots & c_{2K-1}
   \end{array} \right)
   \left(\begin{array}{c} a_1 \\ \vdots \\ a_K \end{array} \right)
 = \left(\begin{array}{c} c_0 \\ \vdots \\ c_{K-1} \end{array} \right) \; .
\label{lin_eq1}
\end{equation}
The second step consists in determining the parameters $z_k$ in 
Eq.~(\ref{c_n:eq}).
Using Eqs.~(\ref{a_k_def}) and (\ref{c_n:eq}) we obtain
\begin{equation}
   c_n=\sum_{k=1}^K a_k c_{n+k}
 = \sum_{l=1}^K \sum_{k=1}^K a_k d_l z_l^{n+k} \; ,
\end{equation}
and thus
\begin{equation}
 \sum_{k=1}^K \left[ \sum_{l=1}^K a_l z_k^{n+l}-z_k^n \right] d_k = 0 \; .
\label{poly1:eq}
\end{equation}
Equation~(\ref{poly1:eq}) is satisfied for arbitrary sets of amplitudes $d_k$
when $z_k$ is a zero of the polynomial
\begin{equation}
 \sum_{l=1}^K a_l z^l - 1 = 0 \; .
\label{poly2:eq}
\end{equation}
The parameters $z_k=\exp{(-{\rm i}w'_k\tau)}$ and thus the frequencies
\begin{equation}
 w'_k = {{\rm i}\over\tau} \log(z_k)
\label{wk:eq}
\end{equation}
are therefore obtained by searching for the zeros of the polynomial 
in Eq.~(\ref{poly2:eq}).
Note that this is the only nonlinear step of the algorithm, and numerical
routines for finding the roots of polynomials are well established.
In the third and final step, the amplitudes $d_k$ are obtained from the
linear set of equations
\begin{equation}
c_n = \sum_{k=1}^K d_k z_k^n \; , \quad n = 0, \dots, K-1 \; .
\label{lin_eq2}
\end{equation}
To summarize, the LP method reduces the {\em nonlinear\/} set of 
equations (\ref{c_n:eq}) for the variational parameters $\{z_k,d_k\}$ to 
two well-known problems, i.e., the solution of two {\em linear\/} 
sets of equations (\ref{lin_eq1}) and (\ref{lin_eq2}) and the root search of a
polynomial, Eq.~(\ref{poly2:eq}), which is a nonlinear but familiar problem.
The matrices in Eqs.~(\ref{lin_eq1}) and (\ref{lin_eq2}) are a Toeplitz and 
Vandermonde matrix, respectively, and special algorithms are known for
the fast solution of such linear systems \cite{NumRec}.
However, when the matrices are ill-conditioned, conventional $LU$
decomposition of the matrices is numerically more stable, and, furthermore,
an iterative improvement of the solution can significantly reduce errors 
arising from numerical rounding.
The roots of polynomials can be found, in principle, by application
of Laguerre's method \cite{NumRec}.
However, it turns out that an alternative method, i.e., the 
diagonalization of the Hessenberg matrix 
\begin{equation}
 {\bf A} =
 \left(\begin{array}{ccccc}
  -{a_{K-1}\over a_K} & -{a_{K-2}\over a_K} & \cdots &
  -{a_{1}\over a_K} & -{a_{0}\over a_K} \\
  1 & 0 & \cdots & 0 & 0 \\
  0 & 1 & \cdots & 0 & 0 \\
  \vdots & \vdots & \ddots & \vdots & \vdots \\
  0 & 0 & \cdots & 1 & 0 
 \end{array} \right) \quad ,
\label{Hesse:eq}
\end{equation}
for which the characteristic polynomial $P(z)=\det[{\bf A}-z{\bf I}]=0$ 
is given by Eq.~(\ref{poly2:eq}) (with $a_0=-1$), is a numerically more 
robust technique for finding the roots of high degree ($K \gtrsim 60$) 
polynomials \cite{NumRec}.

\subsubsection{Pad\'e Approximant}
As an alternative method for solving the nonlinear system (\ref{c_n:eq})
we now propose to apply the method of Pad\'e approximants (PA) to our 
band-limited signal $c_n$.
Let us assume for the moment that the signal points $c_n$ are known up
to infinity, $n=0,1,\dots\infty$.
Interpreting the $c_n$'s as the coefficients of a Maclaurin series in the
variable $z^{-1}$, we can then define the function 
$g(z)=\sum_{n=0}^\infty c_n z^{-n}$.
With Eq.~(\ref{c_n:eq}) and the sum rule for geometric series we obtain
\begin{equation}
   g(z) \equiv \sum_{n=0}^\infty c_n z^{-n}
 = \sum_{k=1}^K d_k \sum_{n=0}^\infty (z_k/z)^n
 = \sum_{k=1}^K {z d_k \over z-z_k} 
 \equiv {P_{K}(z) \over Q_K(z)} \; .
\label{g_Pade:eq}
\end{equation}
The right-hand side of Eq.~(\ref{g_Pade:eq}) is a rational function 
with polynomials of degree $K$ in the numerator and denominator.
Evidently, the parameters $z_k=\exp{(-{\rm i}w'_k\tau)}$ are the
poles of $g(z)$, i.e., the zeros of the polynomial $Q_K(z)$.
The parameters $d_k$ are calculated via the residues of the last two terms 
of (\ref{g_Pade:eq}).
We obtain
\begin{equation}
 d_k = {P_{K}(z_k) \over z_k Q'_K(z_k)} \; ,
\label{dk_pade:eq}
\end{equation}
with the prime indicating the derivative $d/dz$.
Of course, the assumption that the coefficients $c_n$ are known up to
infinity is not fulfilled  and, therefore, the sum on the left-hand side of 
Eq.~(\ref{g_Pade:eq}) cannot be evaluated in practice.
However, the convergence of the sum can be accelerated by application of PA.
Indeed, with PA, knowledge of $2K$ signal points $c_0,\dots,c_{2K-1}$ is 
sufficient for the calculation of the coefficients of the two polynomials
\begin{equation}
 P_{K}(z) = \sum_{k=1}^{K} b_k z^k  \mbox{~~and~~}
 Q_K(z) = \sum_{k=1}^K a_k z^k - 1 \; .
\end{equation}
The coefficients $a_k$, $k=1,\dots,K$ are obtained as solutions of the 
linear set of equations
\[
 c_n = \sum_{k=1}^K a_k c_{n+k} \; ,  \quad n = 0, \dots, K-1 \; ,
\]
which is identical to Eqs.~(\ref{a_k_def}) and (\ref{lin_eq1}) for LP.
Once the $a$'s are known, the coefficients $b_k$ are given by the 
{\em explicit}\ formula
\begin{equation}
 b_k = \sum_{m=0}^{K-k} a_{k+m} c_{m} \; , \quad k = 1, \dots , K \; .
\end{equation}
It should be noted that the different derivations of LP and PA yield the 
same polynomial whose zeros are the $z_k$ parameters, i.e., the $z_k$ 
calculated with both methods exactly agree.
However, LP and PA do differ in the way the amplitudes, $d_k$, are calculated.
It is also important to note that PA is applied here as a method for signal 
processing, i.e., in a different context to that in Sec.\ \ref{method2:sec},
where the Pad\'e approximant is used for the direct summation of the 
periodic orbit terms in semiclassical trace formulae.

\subsubsection{Signal Diagonalization}
In Refs.~\cite{Wal95,Man97b} it has been shown how the problem of solving 
the nonlinear set of equations (\ref{c_n:eq}) can be recast in the form of 
the generalized eigenvalue problem, 
\begin{equation}
{\bf U} \mbox{\boldmath{$B$}}_k = z_k {\bf S} \mbox{\boldmath{$B$}}_k \; .
\label{geneval}
\end{equation}
The elements of the $K\times K$ operator matrix ${\bf U}$ and
overlap matrix ${\bf S}$ depend trivially upon the $c_n$'s \cite{Man97b}:
\begin{equation}
U_{ij} = c_{i+j+1} \; ; \quad S_{ij} = c_{i+j} \; ; \quad i,j=0,\dots,K-1 \; .
\label{matels}
\end{equation}
Note that the operator matrix ${\bf U}$ is the same as in the linear system 
(\ref{lin_eq1}), i.e., the matrix form of Eq.~(\ref{a_k_def}) of LP.
The matrices ${\bf U}$ and ${\bf S}$ in Eq.~(\ref{geneval}) are complex
symmetric (i.e., non-Hermitian), and the eigenvectors 
$\mbox{\boldmath{$B$}}_k$ are orthogonal with respect to the overlap 
matrix ${\bf S}$,
\begin{equation}
   \left(\mbox{\boldmath{$B$}}_k |{\bf S}| \mbox{\boldmath{$B$}}_{k'} \right)
 = N_k \delta_{kk'} \; ,
\label{Nk:eq}
\end{equation}
where the brackets define a complex symmetric inner product $(a|b)=(b|a)$,
i.e., no complex conjugation of either $a$ or $b.$
The overlap matrix ${\bf S}$ is not usually positive definite
and therefore the $N_k$'s are, in general complex, normalization parameters.
An eigenvector $\mbox{\boldmath{$B$}}_k$ cannot be normalized for $N_k=0$.
The amplitudes $d_k$ in Eq.~(\ref{c_n:eq}) are obtained from the eigenvectors
$\mbox{\boldmath{$B$}}_k$ via
\begin{equation}
 d_k = {1\over{N_k}}
 \left[ \sum_{n=0}^{K-1} c_n \mbox{\boldmath{$B$}}_{k,n} \right]^2 \; .
\label{dk_dsd:eq}
\end{equation}
The parameters $z_k$ in Eq.~(\ref{c_n:eq}) are given as the eigenvalues of 
the generalized eigenvalue problem (\ref{geneval}), and are simply related to 
the frequencies $w'_k$ in Eq.~(\ref{C_bld}) via $z_k=\exp(-{\rm i}w'_k\tau)$.

\bigskip\noindent
The three methods introduced above (LP, PA and SD) look technically
quite different.
With LP the coefficients of the characteristic polynomial (\ref{poly2:eq})
and the amplitudes $d_k$ are obtained by solving two linear sets of
equations (\ref{lin_eq1}) and (\ref{lin_eq2}).
Note that the complete set of zeros $z_k$ of Eq.~(\ref{poly2:eq}) is required
to solve for the $d_k$ in Eq.~(\ref{lin_eq2}).
The PA method is even simpler, as only one linear system, 
Eq.~(\ref{lin_eq1}), has to be solved to determine the coefficients
of the rational function $P_{K}(z)/Q_K(z)$.
Finding the zeros of Eq.~(\ref{poly2:eq}) provides knowledge about selected 
parameters $z_k$, and allows one to calculate the corresponding amplitudes 
$d_k$ via Eq.~(\ref{dk_pade:eq}).
The SD method requires the most numerical effort, because the solution
of the generalized eigenvalue problem (\ref{geneval}) for both the 
eigenvalues $z_k$ and eigenvectors $\mbox{\boldmath{$B$}}_k$ is needed.

It is important to note that the three techniques, in spite of their
different derivations, are mathematically equivalent and provide the same 
results for the parameters $\{z_k,d_k\}$, when the following two conditions 
are fulfilled:
the nonlinear set of equations (\ref{c_n:eq}) has a unique solution, when,
firstly, the matrices ${\bf U}$ and ${\bf S}$ in Eq.~(\ref{matels})
have a non-vanishing determinant ($\det{\bf U}\ne 0$, $\det{\bf S}\ne 0$), 
and, secondly, the parameters $z_k$ are non-degenerate 
($z_k\ne z_{k'}$ for $k\ne k'$).
These conditions guarantee the existence of a unique solution of the linear 
equations (\ref{lin_eq1}) and (\ref{lin_eq2}), the non-singularity of the 
generalized eigenvalue problem (\ref{geneval}), and the non-vanishing of both 
the derivatives $Q'_K(z_k)$ in Eq.~(\ref{dk_pade:eq}) and the normalization 
constants $N_k$ in Eqs.~(\ref{Nk:eq}) and (\ref{dk_dsd:eq}).
Equation~(\ref{c_n:eq}) usually has no solution in the case of degenerate 
$z_k$ parameters, however, degeneracies can be handled with a generalization 
of the ansatz (\ref{c_n:eq}) and modified equations for the calculation of 
the parameters.
Here, we will not further discuss this special case.

While the parameters $z_k$ in Eq.~(\ref{c_n:eq}) are usually unique, the
calculation of the frequencies $w'_k$ via Eq.~(\ref{wk:eq}) is not unique,
because of the multivalued property of the complex logarithm.
To obtain the ``correct'' frequencies it is necessary to appropriately 
adjust the range $\Delta w$ of the frequency filter and the step width 
$\tau$ of the band-limited signal (\ref{C_bld}).
From our numerical experience we recommend the following procedure.
The most convenient approach is to choose first the center $w_0$ of the 
frequency window and the number $K$ of frequencies within that window.
Note that $K$ determines the dimension of the linear systems, and hence 
the degree of the polynomials which have to be handled numerically, and is 
therefore directly related to the computational effort required.
Frequency windows are selected to be sufficiently narrow to yield 
values for the rank between $K\approx 50$ and $K\approx 200$.
The step width for the band-limited signal should be chosen as 
\begin{equation}
 \tau = {s_{\rm max}\over 2K} \; ,
\end{equation}
with $s_{\rm max}$ being the total length of the periodic orbit signal.
The relation $z_k=\exp{(-{\rm i}w'_k\tau)}$ projects the frequency window
$w'\in [-\Delta w,+\Delta w]$ onto the unit circle in the complex plane
when the range of the frequency window is chosen as 
\begin{equation}
 \Delta w = {\pi\over\tau} = {2\pi K\over s_{\rm max}} \; .
\end{equation}
When calculating the complex logarithm with $\arg\log z\in [-\pi,+\pi]$, 
Eq.~(\ref{wk:eq}) provides the ``correct'' shifted frequencies $w'_k$ and 
thus the frequencies $w_k=w_0+w'_k$.

To achieve convergence, the length $s_{\rm max}$ of the periodic orbit signal 
must be sufficiently long to ensure that the number of semiclassical 
eigenvalues within the frequency window is less than $K$.
As a consequence the harmonic inversion procedure usually provides not only
the true semiclassical eigenvalues but also some spurious resonances.
The spurious resonances are identified by low or near zero values of the
corresponding amplitudes $d_k$ and can also be detected by analyzing
the shifted band-limited signal, i.e., signal points $c_1,\dots,c_{2K}$
instead of $c_0,\dots,c_{2K-1}$.
The true frequencies usually agree to very high precision, while spurious
frequencies show by orders of magnitude larger deviations.

The harmonic inversion method introduced above will be applied in Sec.\ 
\ref{3disk:subsec} to the periodic orbit quantization of the three-disk 
scattering system, and the semiclassical resonances will be compared with 
results obtained by the cycle-expansion technique \cite{Cvi89,Eck95,Wir99}
and the direct application of the Pad\'e approximant to periodic orbit sums
discussed in the next Section \ref{method2:sec}.

\section{Semiclassical quantization by Pad\'e approximants to periodic
orbit sums}
\label{method2:sec}
The method presented in the previous Section is based on signal processing
(harmonic inversion) of the periodic orbit signal.
In this Section we introduce our second method for semiclassical
quantization, based on the direct application of the Pad\'e approximant 
to periodic orbit sums.

The PA to a complex function $f(z)$ is defined as a ratio of two polynomials 
and can be computed from the coefficients $a_n$ of the Maclaurin expansion 
of $f(z)$, i.e., a power series
\[
 f(z)=\sum_{n=0}^\infty a_nz^n
\]
with finite or even zero radius of convergence in $z$ \cite{Bel89}.
However, Eq.\ (\ref{gE_sc}) does not have the functional form of a
Maclaurin power series expansion of $g(E)$ in the energy $E$.
Even disregarding this limitation, a direct computation of the PA to the 
sum in Eq.\ (\ref{gE_sc}) would be numerically unstable due to the 
typically large number of periodic orbit terms. Nevertheless,
considering $E$ as a parameter, $g(E)$ can be rearranged and written as 
a formal power series in an auxiliary variable $z$,
\begin{equation}
 g(z;E) = \sum_n \sum_{\mu_{\rm po}=n}
   {\cal A}_{\rm po}(E){\rm e}^{{\rm i}S_{\rm po}(E)/\hbar} z^n
\; \equiv \; \sum_n a_n(E)z^n \; ,
\label{g_z_def:eq}
\end{equation}
where the maximum value of $n$ required for convergence of the PA is 
relatively small compared with the number of periodic orbit terms.
Of course, the arrangement (\ref{g_z_def:eq}) of the periodic orbit sum as a
power series is not unique, and expansions similar to Eq.~(\ref{g_z_def:eq})
can be used for other ordering parameters $n$ of the orbits, e.g., 
the cycle length in systems with a symbolic code.
However, if no symbolic dynamics exists, the sorting of orbits by their 
Maslov index is natural both physically and as a way to introduce an integer 
summation index and will be justified in Sec.\ \ref{results:sec} by the 
successful numerical application of the method to a regular system without 
symbolic dynamics.
Note that the PA to the periodic orbit sum cannot be applied
without any ordering parameter, i.e., when no symbolic dynamics 
exists and all Maslov indices are zero, which is the case, e.g.,
for the Riemann zeta function as a mathematical model for periodic orbit 
quantization \cite{Mai98a}.
The true value $g(E)$ of the periodic orbit sum is obtained by setting 
$z=1$ in Eq.~(\ref{g_z_def:eq}), i.e., $g(E)=g(1;E)$. 
In such a case, we have a point PA which is given as a ratio of two 
polynomials in $z$ whose coefficients are non-polynomial functions of $E$ 
all at a fixed value of $z$.
The usual implementation of the PA as a ratio of two polynomials in $z$
whose coefficients are computed, e.g., via the Longman algorithm \cite{Lon79}
would be advantageous if $g(z;E)$ were required for many values of $z$.
In the present case, at each given energy $E$ only one fixed value $z=1$ is 
needed and the PA is most efficiently computed by means of the recursive 
Wynn $\varepsilon$-algorithm \cite{Wyn56}. 

To briefly describe the $\varepsilon$-algorithm, we introduce a sequence of 
partial sums $\{A_n\}$ which converges to (or diverges from) its limit $A$ 
as $n\to\infty$. In the case of divergence, $A$ is called the `anti-limit' of
$\{A_n\}\,,$ as $n\to\infty$. Further, let $F$ be a transformation which maps 
$\{A_n\}$ into another sequence $\{B_n\}$.
The mapping $F$ will represent an accelerator, i.e., the sequence $\{B_n\}$ 
will converge to the same limit $A$ faster than $\{A_n\}$ if the condition 
\[
 {B_n-B \over A_n-A} \to 0
\]
is fulfilled as $n\to\infty$.
In addition, the same $F$ can be applied to wildly divergent sequences 
$\{A_n\}$. Only nonlinear mappings can simultaneously accomplish both goals 
to accelerate slowly convergent and induce convergence into divergent 
sequences. The transformation $F$ will be nonlinear if its coefficients 
depend on $A_n$, e.g., the so-called $e$-algorithm of Shanks \cite{Sha55}, 
whose mapping $F$ is the operator $e_k$ which converts the sequence 
$\{A_n\}$ into $\{B_n\}$ via 
\begin{equation}
 e_k(A_n)=B_{k,n}=[n+k/k]
\end{equation}
with $n\ge 0$ and $n\ge k$.
This is the well-known Aitken $\Delta^2$-iteration process 
(i.e., the simplest PA, [1/1]) extended to higher orders $k$. 
The general term in the $k$th-order transform $B_{k,n}$ of $A_n$ can be 
computed efficiently from the stable and recursive $\varepsilon$-algorithm 
of Wynn \cite{Wyn56}, viz., 
\begin{equation}
 e_s(A_m)=\varepsilon_{2s}^{(m)}=[m+s/s] \; ,
\end{equation}
where
\begin{equation}
   \varepsilon_{s+1}^{(m)}
 = \varepsilon_{s-1}^{(m+1)}
 + {1\over \varepsilon_s^{(m+1)}-\varepsilon_s^{(m)}} \quad ; \quad
   m, s \ge 0
\label{eps_alg:eq}
\end{equation}
with $\varepsilon_{-1}^{(m)}=0$, $\varepsilon_0^{(m)}=A_m$,
$\varepsilon_{2s+1}^{(m)}=1/e_s(\Delta A_m)$, and where $\Delta$ 
is the forward difference operator: $\Delta x_j=x_{j+1}-x_j$.
When $\{A_n\}$ is the sequence of partial sums of a power series, the
two-dimensional array $\varepsilon_{2s}^{(m-s)}$ yields the upper half of the
well known Pad\'e table $[m/s]$. However, the $\varepsilon$-algorithm
need not necessarily be limited to power series.

The procedure to apply the above PA to semiclassical quantization 
by summation of periodic orbit terms works as follows.
For a given system we calculate the periodic orbits up to a chosen maximum
ordering parameter $n \le n_{\rm max}$, where $n$ can be but is not
necessarily related to the Maslov indices of orbits.
Note that this set of orbits usually differs from the set of orbits
with classical action $S_{\rm po}\le S_{\rm max}$, which is required for
periodic orbit quantization by harmonic inversion in Sec.\ \ref{method1:sec}.
From the periodic orbit amplitudes ${\cal A}_{\rm po}$ (including the
phases $\exp(-{\rm i}{\pi\over 2}\mu_{\rm po})$ given by the Maslov indices)
and actions $S_{\rm po}$, we compute the partial sums 
(e.g., with $n\le \mu_{\rm max}$ the Maslov index):
\begin{equation}
 A_n = \sum_{\mu_{\rm po}\le n} {\cal A}_{\rm po}(E)
   {\rm e}^{{\rm i}S_{\rm po}(E)/\hbar} \; .
\label{A_n:eq}
\end{equation}
The sequence $\{A_n\}$ of partial sums is used as input to the
$\varepsilon$-algorithm, Eq.\ (\ref{eps_alg:eq}), to obtain
a converged value $g(E)$ of the periodic orbit sum (\ref{gE_sc}).
The semiclassical eigenenergies or resonances are given as the poles 
of $g(E)$ and are obtained by searching numerically  for the zeros
of the reciprocal function $1/g(E)$. 
Such a search requires the evaluation of, e.g., ${\cal A}_{\rm {po}}(E)$ 
and $S_{\rm {po}}(E)$, at complex values of $E$.
This can be done in a straightforward manner for the scaling systems 
considered in this paper.
The number of the poles of $g(E)$ is not constrained by the size of 
the sequence $\{A_n\}$ of partial sums since our PA is a ratio of two 
non-polynomial functions of $E$.

It is important to note that the Pad\'e approximant is applied here in a 
completely different context than in the first method where it is used as 
a tool for signal processing.
In Sec.\ \ref{method1:sec} we have shown that the three techniques
LP, PA and SD used for the harmonic inversion of band-limited signals 
are mathematically equivalent.
However, the two methods for periodic orbit quantization introduced in
Sections \ref{method1:sec} and \ref{method2:sec} are not necessarily 
equivalent, and it may well be that one or the other method is more 
appropriate for the semiclassical quantization of a given physical system.
Numerical results for two physical systems with completely different 
dynamical properties will be presented in the next Sec.\ \ref{results:sec}.

\section{Results and discussion}
\label{results:sec}
In this section we want to demonstrate the efficiency and accuracy of
the methods introduced above by way of two examples:
the three-disk repeller as an open physical system with classically
chaotic dynamics and the circle billiard as a bound regular system.
Both systems have previously been investigated by means of FD 
\cite{Mai97a,Mai98a,Mai99c,Mai98b,Mai99e,Wei00}.
The three-disk repeller has also served as a prototype for the development 
and application of cycle expansion techniques \cite{Cvi89,Eck95,Wir99}, 
and for this system we will compare the convergence properties of
three different methods, viz.\ harmonic inversion, Pad\'e approximant 
and cycle expansion.

\subsection{The three-disk repeller}
\label{3disk:subsec}
As the first example we consider a billiard system consisting of three
identical hard disks with radius $R$, displaced from each other
by the same distance $d$.
This simple, albeit nontrivial, scattering system has served as a
model for the development of the cycle expansion method 
\cite{Cvi89,Eck93,Eck95,Wir99} and periodic orbit quantization by
harmonic inversion \cite{Mai97a,Mai98a,Mai99c}.
The three-disk scattering system is invariant under the symmetry operations
of the group $C_{3v}$, {\it i.e.\/}, three reflections at symmetry lines 
and two rotations by $2\pi/3$ and $4\pi/3$.
Resonances belong to one of the three irreducible subspaces $A_1$, $A_2$, 
and $E$ \cite{Cvi93a}.
As in most previous work we concentrate on the resonances of the 
subspace $A_1$ for the three-disk repeller with $R=1$ and $d=6$.
In billiards, which are scaling systems, the shape of periodic orbits
does not depend on the energy $E$, and the classical action is given by 
the length $L$ of the orbit ($S_{\rm po}=\hbar kL_{\rm po}$), where 
$k=|{\bf k}|=\sqrt{2ME}/\hbar$ is the absolute value of the wave vector 
to be quantized.
We have calculated all periodic orbits with Maslov index $\mu_{\rm po}\le 30$,
which corresponds to the set of orbits with cycle length $n\le 15$.

\subsubsection{Harmonic inversion of the periodic orbit signal}
\begin{figure*}
\vspace{18.5cm}
\includegraphics{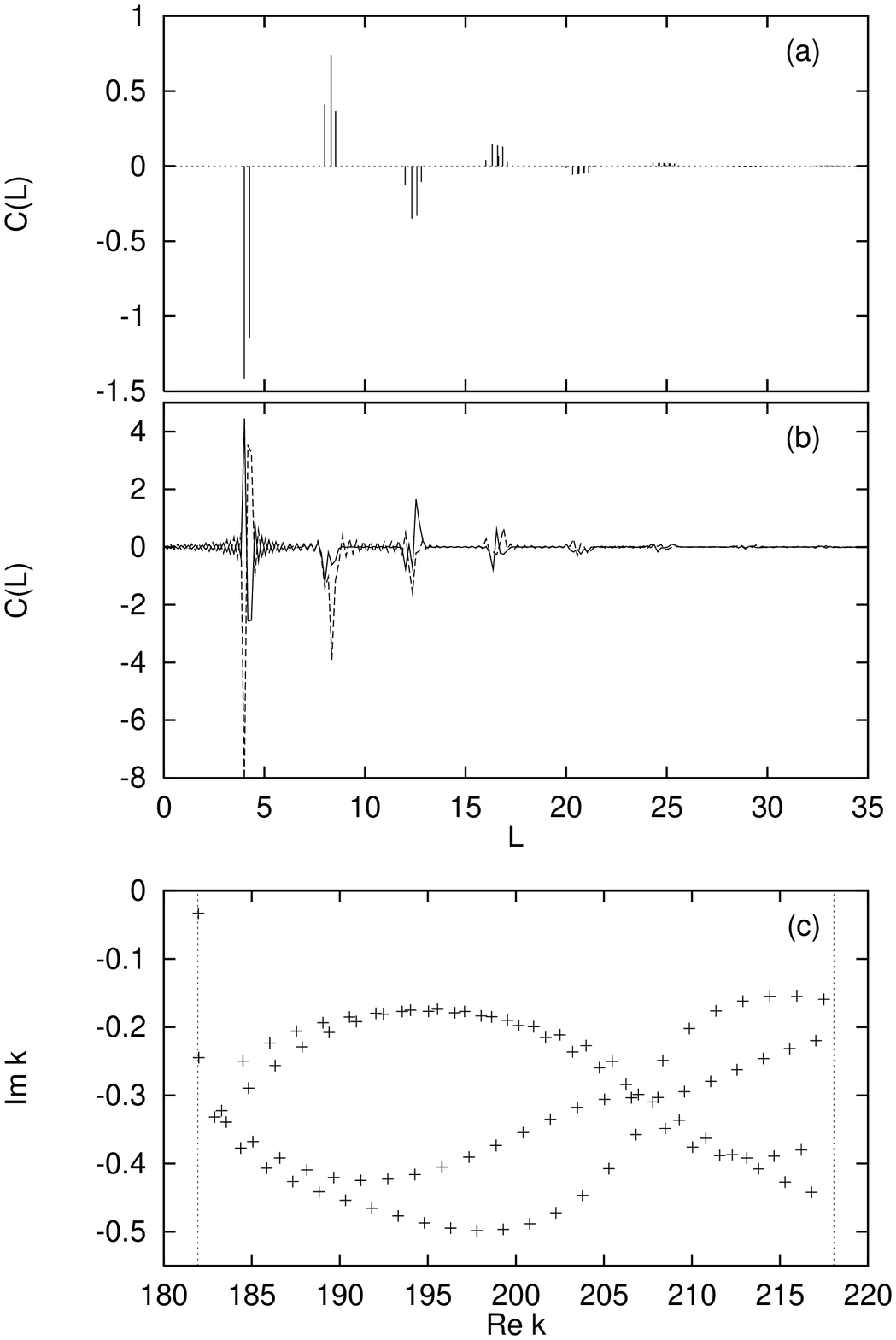}
\caption{(a) Periodic orbit recurrence signal for the three-disk scattering 
system with $R=1$, $d=6$ without filtering.
The signal in the region $L\le 35$ consists of 93 non-equidistant 
periodic orbit contributions (including multiple repetitions).
(b) Same as (a) filtered with frequency window $w\in [182, 218]$.
The band-limited signal consists of 201 equidistant data points with 
$\Delta L=0.175$.
The solid and dashed lines are the real and imaginary part of $C(L)$,
respectively.
(c) Semiclassical resonances obtained by harmonic inversion of the
band-limited signal $C(L)$ in (b).
The dotted lines mark the borders of the frequency window.
}
\label{fig1}
\end{figure*}

We first calculate the semiclassical resonances of the three-disk repeller
by the method introduced in Sec.\ \ref{method1:sec}, viz.\ harmonic inversion
of the periodic orbit signal [see Eq.~(\ref{C_sc})]
\[
 C^{\rm sc}(L) = \sum_{\rm po} {\cal A}_{\rm po} \delta(L-L_{\rm po}) \; .
\]
[Setting $\hbar=1$, we use $s=L$ in what follows.]
In Fig.\ \ref{fig1}a we present the periodic orbit signal $C^{\rm sc}(L)$
for the three-disk repeller in the region $0\le L\le L_{\rm max}=35$.
The signal is given as a periodic orbit sum of delta functions
$\delta(L-L_{\rm po})$ weighted with the periodic orbit amplitudes
${\cal A}_{\rm po}$.
The groups with oscillating sign belong to periodic orbits with adjacent
cycle lengths.
Signals of this type have been analyzed (after convolution with a narrow
Gaussian function, see Eq.~(\ref{C_sc_sigma})) by FD
in Refs.~\cite{Mai97a,Mai98a,Mai99a,Mai99b,Mai99c}.
We now illustrate the harmonic inversion of band-limited periodic orbit 
signals by LP, PA and SD.

In a first step, a band-limited periodic orbit signal is constructed
as described in Sec.\ \ref{analyt_dec:sec}.
As an example we choose $K=100$ as the rank of the nonlinear set of 
equations (\ref{C_bld}), and $k_0=200$ as the center of the frequency window.
The width of the frequency window is given by 
$\Delta k=2\pi K/L_{\rm max}=200\pi/35\approx 18.0$.
The step width of the band-limited signal is 
$\tau=\Delta L=L_{\rm max}/2K=0.175$.
The signal points $c_n=C^{\rm sc}_{\rm bl}(L=n\Delta L)$, with 
$n=0,\dots,2K$ are calculated with the help of Eq.~(\ref{C_sc_bl}) 
and presented in Fig.\ \ref{fig1}b.
The solid and dashed lines are the real and imaginary parts of 
$C^{\rm sc}_{\rm bl}(n\Delta L)$, respectively.
The modulations with spacings $\pi/\Delta k$ result from the superposition
of the sinc-like functions in Eq.~(\ref{C_sc_bl}).

The band-limited periodic orbit signal $C^{\rm sc}_{\rm bl}(n\Delta L)$
can now be analyzed, in a second step, with one of the harmonic inversion
techniques introduced in Sec.\ \ref{hi:sec}, viz.\ LP, PA or SD.
The resonances obtained by LP are presented as plus symbols in Fig.\ 
\ref{fig1}c.
The dotted lines at ${\rm Re}~k=182$ and ${\rm Re}~k=218$ show the borders
of the frequency window.
The two symbols very close to the border on the left indicate spurious 
resonances.

A long range spectrum can be obtained by choosing several values $w_0$ for
the center of the frequency window in such a way that the windows slightly
overlap.
By analyzing a periodic orbit signal similar to that in Fig.\ \ref{fig1}b 
but with an increased signal length, $L_{\rm max}=55$ we have calculated 
the semiclassical resonances of the three-disk repeller in the range 
$0\le {\rm Re}~k\le 250$.
It turns out that they are even more accurate than those obtained 
previously \cite{Mai98a,Mai99c} using FD.
For more details see Ref.~\cite{Mai00}.
Part of the resonances in the range $25\le {\rm Re}~k\le 65$ are marked
as squares in Fig.~\ref{fig2}.
The comparison of resonances in Fig.~\ref{fig2} obtained by various 
semiclassical quantization methods will be discussed below.

\subsubsection{Pad\'e approximant to the periodic orbit sum}
We now apply our second method introduced in Sec.\ \ref{method2:sec}, 
the PA to the periodic orbit sum to the same system as discussed above, 
viz.\ the three-disk repeller with $R=1$ and $d=6$.
We have calculated the partial sums $\{A_n\}$ of periodic orbit terms 
(see Eq.\ \ref{A_n:eq}) using all periodic orbits with cycle length 
$n\le 15$, which corresponds to the set of orbits with Maslov index 
$\mu_{\rm po}\le 30$.
The sequence of the partial sums $\{A_n\}$ of periodic orbit 
terms in Eq.\ (\ref{A_n:eq}) converges for wave numbers $k$ above the
borderline ${\rm Im}~k = -0.121\,557$ \cite{Cvi89} which separates the 
domain of absolute convergence of the periodic orbit sum from the domain 
where analytic continuation is necessary, but strongly diverges deep in 
the complex plane, where the resonance poles are located.
This is illustrated in Fig.~\ref{fig3} for two different wave numbers $k$.
The dashed line and the plus symbols in Fig.~\ref{fig3}a show the
convergence of the sequence $\{A_n\}$ at $k=150-0.1{\rm i}$.
What is plotted is the absolute value of the difference between two
consecutive terms, $\varepsilon=|A_n-A_{n-1}|$.
As can be seen this sequence is slowly convergent, and about two to 
three significant digits are obtained at $n=15$.
The convergence can be accelerated using the PA, as is seen by the 
solid line and squares in Fig.~\ref{fig3}a showing the error values 
$\varepsilon=|A_n^{\rm PA}-A_{n-1}^{\rm PA}|$ for the sequence of the 
Pad\'e approximants $\{A_n^{\rm PA}\}$ to the periodic orbit sum.
The periodic orbit sum has converged to six significant digits already for
$n=10$.
The situation is even much more dramatic in the deep complex plane
(${\rm Im}~k<-0.122$), {\it e.g.\/}, at $k=150-0.5{\rm i}$.
Here, the sequence of the partial sums $\{A_n\}$ of periodic orbit terms 
exhibits {\em exponential} divergence, as can be seen from the dashed line
and plus symbols in Fig.~\ref{fig3}b.
Nevertheless, this sequence converges when subjected to the PA implemented
through the $\varepsilon$-algorithm (solid line and squares 
in Fig.~\ref{fig3}b).

The resonances of the three-disk scattering systems have been obtained
by a numerical two-dimensional root search in the complex $k$-plane for 
the zeros of the function $1/g(k)$, where $g(k)$ is the PA to the periodic
orbit sum.
The subset of the semiclassical resonances in the range 
$25\le {\rm Re}~k\le 65$ are marked by the plus symbols in Fig.~\ref{fig2}.
They agree well with the squares obtained by harmonic inversion of the
periodic orbit signal.
For the resonances shown in Fig.~\ref{fig2} we will now discuss the
applicability and limitations of the different semiclassical methods 
to some more extent.

\subsubsection{Harmonic inversion and Pad\'e approximant vs.\ cycle expansion}
\label{compare:subsubsec}
The three-disk scattering system discussed above has purely hyperbolic 
classical dynamics and has been used extensively as the prototype model 
for the cycle expansion techniques \cite{Cvi89,Eck93,Eck95,Wir99}.
As has been shown by Voros \cite{Vor88}, Gutzwiller's trace formula for
unstable periodic orbits can be recast in the form of an infinite and 
non-convergent Euler product over all periodic orbits.
When the periodic orbits obey a symbolic dynamics the semiclassical 
eigenenergies or resonances can be obtained as the zeros of the cycle
expanded Gutzwiller-Voros zeta function.
Unfortunately, the convergence of the cycle expansion is restricted, 
due to poles of the Gutzwiller-Voros zeta function \cite{Eck93}.
The domain of analyticity of semiclassical zeta functions can be extended 
\cite{Cvi93b,Cvi93c} resulting in the ``quasiclassical zeta function'' 
\cite{Cvi93c,Wir99}, which is an entire function for the three-disk repeller.
This approach allows one to calculate semiclassical resonances in critical 
regions where the Gutzwiller-Voros zeta function does not converge, at the 
cost, however, of many extra spurious resonances and with the rate of 
convergence being slowed down tremendously \cite{Wir99}.

It is very interesting and illustrative to compare the convergence properties
of the various methods which have been applied for periodic orbit
quantization of the three-disk repeller, viz.\ the two methods introduced 
in Sections \ref{method1:sec} and \ref{method2:sec} of this paper and the 
cycle expansion technique.
We will demonstrate that the harmonic inversion method and the PA to periodic
orbit sums provide semiclassical resonances in energy regions where the 
cycle expansion of the Gutzwiller-Voros zeta function does not converge.
With the limited numerical accuracy of harmonic inversion by FD applied 
in Ref.~\cite{Mai98a}, the semiclassical resonances of the three-disk 
repeller in the region ${\rm Im}~k<-0.6$ were somewhat unreliable.
The analysis of band-limited periodic orbit signals introduced in the 
present paper now allows us to calculate semiclassical resonances of much
improved accuracy even deep in the complex plane.

\begin{figure*}
\vspace{18.5cm}
\includegraphics{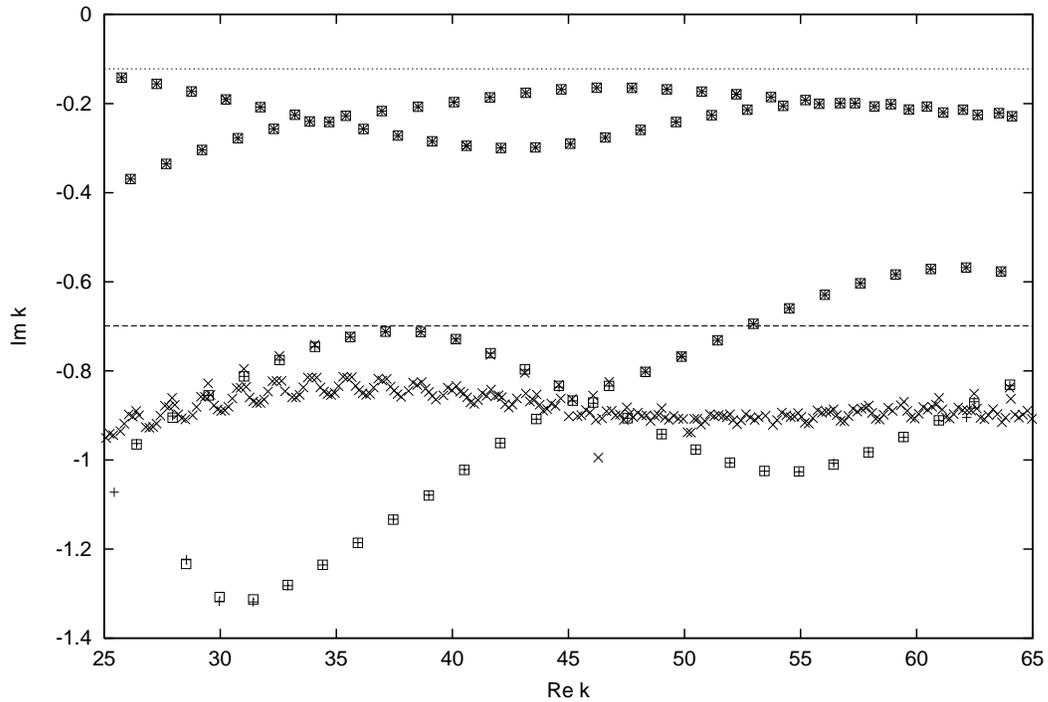}
\caption{
Semiclassical resonances ($A_1$ subspace) for the three-disk scattering  
system with $R=1$, $d=6$.
Squares: Harmonic inversion of the semiclassical recurrence signal;
Plus symbols: Pad\'e approximant to the periodic orbit sum;
Crosses: Cycle expansion of the Gutzwiller-Voros zeta function
\protect\cite{Wir99}.
The dotted and dashed lines mark the borderline for absolute convergence
of Gutzwiller's trace formula $({\rm Im}~k=-0.121\,557)$ and the
Gutzwiller-Voros zeta function $({\rm Im}~k=-0.699\,110)$, respectively.
The Pad\'e approximant and harmonic inversion results are found to 
converge deeper in the complex plane than do the Gutzwiller-Voros 
zeta function results.
}
\label{fig2}
\end{figure*}

\begin{figure}
\vspace{18.5cm}
\includegraphics{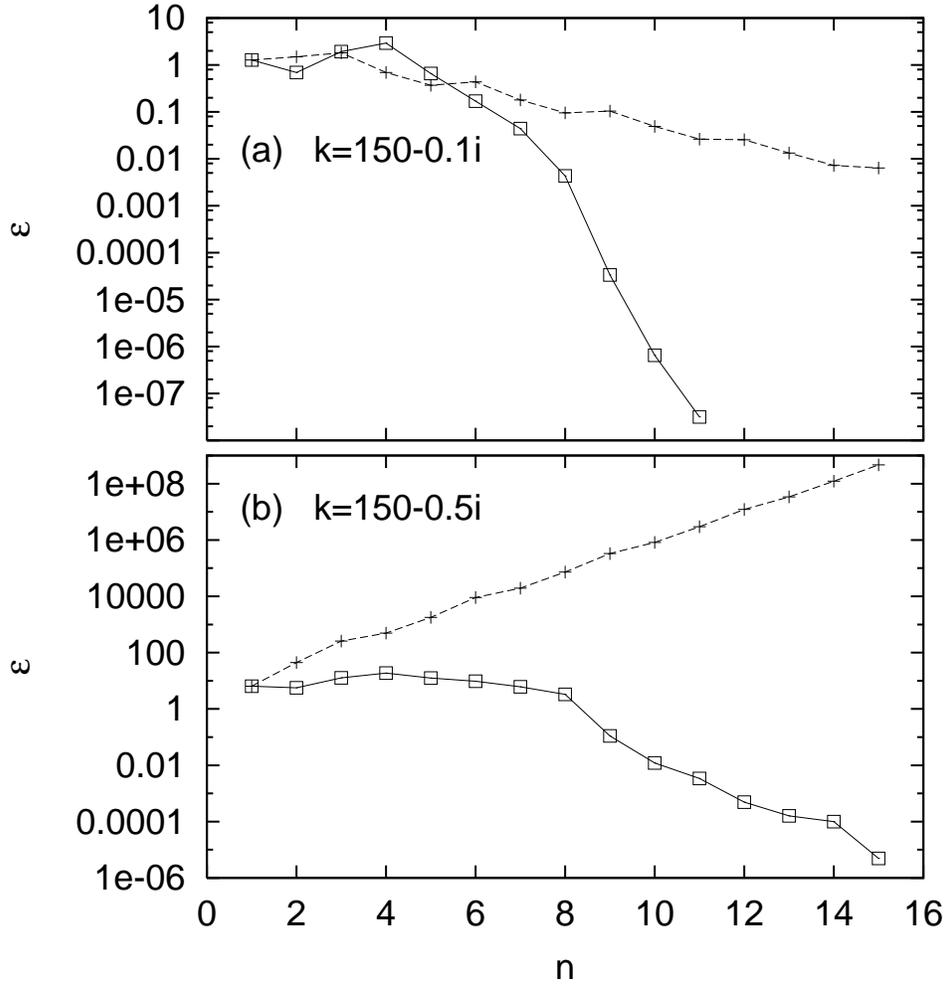}
\caption{
Convergence vs.\ exponential divergence of the partial periodic orbit 
sums for the three-disk scattering system with $R=1$, $d=6$ as functions 
of the order $n$ at (a) complex wave number $k=150-0.1{\rm i}$ and 
(b) $k=150-0.5{\rm i}$.
Dashed lines and plus symbols: 
Error values $\varepsilon=|A_n-A_{n-1}|$ for the sequence $A_n$ 
without Pad\'e approximation.
Solid lines and squares: 
Error values $\varepsilon=|A_n^{\rm PA}-A_{n-1}^{\rm PA}|$ for the Pad\'e 
approximant to the periodic orbit sums.
}
\label{fig3}
\end{figure}

Fig.~\ref{fig2} presents the results of the three different methods in the 
region $25 \le {\rm Re}~k \le 65$.
The resonances calculated by harmonic inversion and the PA to periodic orbit
sum are marked by the squares and plus symbols as already mentioned above.
The crosses label the resonances obtained by the cycle expansion of the 
Gutzwiller-Voros zeta function \cite{Wir99}.
The dotted line in Fig.~\ref{fig2} indicates the borderline, 
${\rm Im}~k=-0.121\,557$ \cite{Cvi89}, which separates the domain of 
absolute convergence of Gutzwiller's trace formula from the region where 
analytic continuation is necessary.
For the two resonance bands slightly below this border the results of all
three semiclassical quantization methods are in perfect agreement.
The dashed line in Fig.~\ref{fig2} marks the borderline of 
absolute convergence of the Gutzwiller-Voros zeta function,
at ${\rm Im}~k=-0.699\,110$ \cite{Cvi93b}.
The Gutzwiller-Voros zeta function provides several spurious resonances
which accumulate at ${\rm Im}~k\approx -0.9$, i.e., slightly below 
the borderline of absolute convergence (see the crosses in Fig.\ \ref{fig2}).
The resonances in the region ${\rm Im}~k<-0.9$, especially those 
belonging to the fourth band, are not described by the Gutzwiller-Voros 
zeta function but are obtained by both the harmonic inversion method and 
the Pad\'e approximant to the periodic orbit sum 
(see the squares and plus symbols in Fig.~\ref{fig2}, respectively).

\subsection{The circle billiard}
As the second example we now demonstrate the applicability of our
semiclassical quantization methods to the circle billiard.
This system has previously been investigated as a model for periodic
orbit quantization of integrable systems \cite{Bal72,Rei96}, however,
to the best of our knowledge, has not yet been treated by the cycle 
expansion technique \cite{Cvi89} or pseudo-orbit expansion \cite{Ber90}.

The exact quantum mechanical eigenvalues $E=\hbar^2k^2/2M$ of the circle
billiard are given by zeros of Bessel functions $J_{|m|}(kR)=0$, where 
$m =0,\pm 1,\pm2,,\dots$ is the angular momentum quantum number and $R$ 
is the radius of the circle.
The semiclassical eigenvalues can be obtained by an Einstein-Brillouin-Keller
(EBK) torus quantization \cite{Per77} resulting in the quantization 
condition
\begin{equation}
 kR\sqrt{1-(m/kR)^2} - |m|\arccos{|m|\over kR} = \pi\left(n+{3\over 4}\right)
\label{EBK}
\end{equation}
where $n=0,1,2,\dots$ is the radial quantum number.
In the following we choose $R=1$.
States with angular momentum quantum number $m\ne 0$ are twofold degenerate.

\begin{figure}
\vspace{18.5cm}
\includegraphics{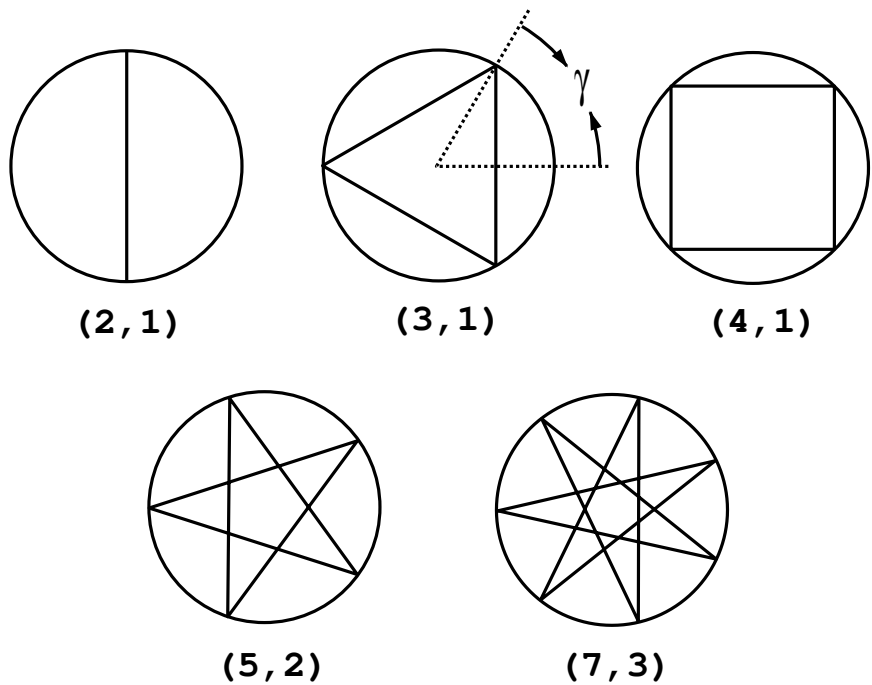}
\caption{Some examples of periodic orbits of the circle billiard.
The orbits are labeled by the numbers $(m_r,m_\varphi)$ which correspond 
to the number of sides of the polygons and the number of turns around the 
center.  The angle $\gamma$ is given by $\gamma=\pi m_\varphi/m_r$.
\label{fig4}
}
\end{figure}

The periodic orbits of the circle billiard have the form of regular polygons.
They can be labeled by two integer numbers $m_r$ and $m_\varphi$ with
the restriction $m_r \ge 2m_\varphi$ which can be shown to be identical 
with the number of sides of the corresponding polygon and its number of 
turns around the center of the circle, respectively \cite{Bal72}. 
Some examples are given in Fig.~\ref{fig4}.
The angle between two bounces is a rational multiple of $2\pi$, i.e.,
the periods $L_{\rm po}$ are obtained from the condition
\begin{equation}
 L_{\rm po} = 2m_r \sin \gamma \; ,
\label{L_circ:eq}
\end{equation}
with $\gamma\equiv\pi m_\phi/m_r$. 
Periodic orbits with $m_r\ne 2m_\phi$ can be traversed in two directions
and thus have multiplicity 2.
For the amplitudes ${\cal A}_{\rm po}$ of the circle billiard, the
Berry-Tabor formula \cite{Ber76} for integrable systems yields
\begin{equation}
   {\cal A}_{\rm po}
 = \sqrt{\pi\over 2}{L_{\rm po}^{3/2}\over m_r^2} \,
   {\rm e}^{-{\rm i}{\pi\over 2}\left(3m_r+{1\over 2}\right)} \; ,
\label{A_circ:eq}
\end{equation}
with $\mu_{\rm po}=3m_r$ being the Maslov index.

The periodic orbit quantities can be used to set up the semiclassical
recurrence signal (\ref{C_sc}) for the circle billiard which can then
be analyzed by harmonic inversion to extract the eigenenergies.
Detailed comparisons between results obtained by harmonic inversion 
and the EBK torus quantization (\ref{EBK}) are presented in 
Refs.\ \cite{Mai99c,Mai98b,Mai99e,Wei00} and show excellent agreement.
The harmonic inversion method can even be generalized, firstly, to 
the harmonic inversion of cross-correlated periodic orbit sums
\cite{Mai99e,Hor00} which allows to significantly reduce the number of 
orbits required for semiclassical quantization, and, secondly, to
include higher order $\hbar$ corrections in the periodic orbit sum.
We do not report these results here but refer the reader to the literature
for details.
In what follows we want to apply our second method introduced in Sec.\
\ref{method2:sec} to the circle billiard.

\begin{figure}
\vspace{18.5cm}
\includegraphics{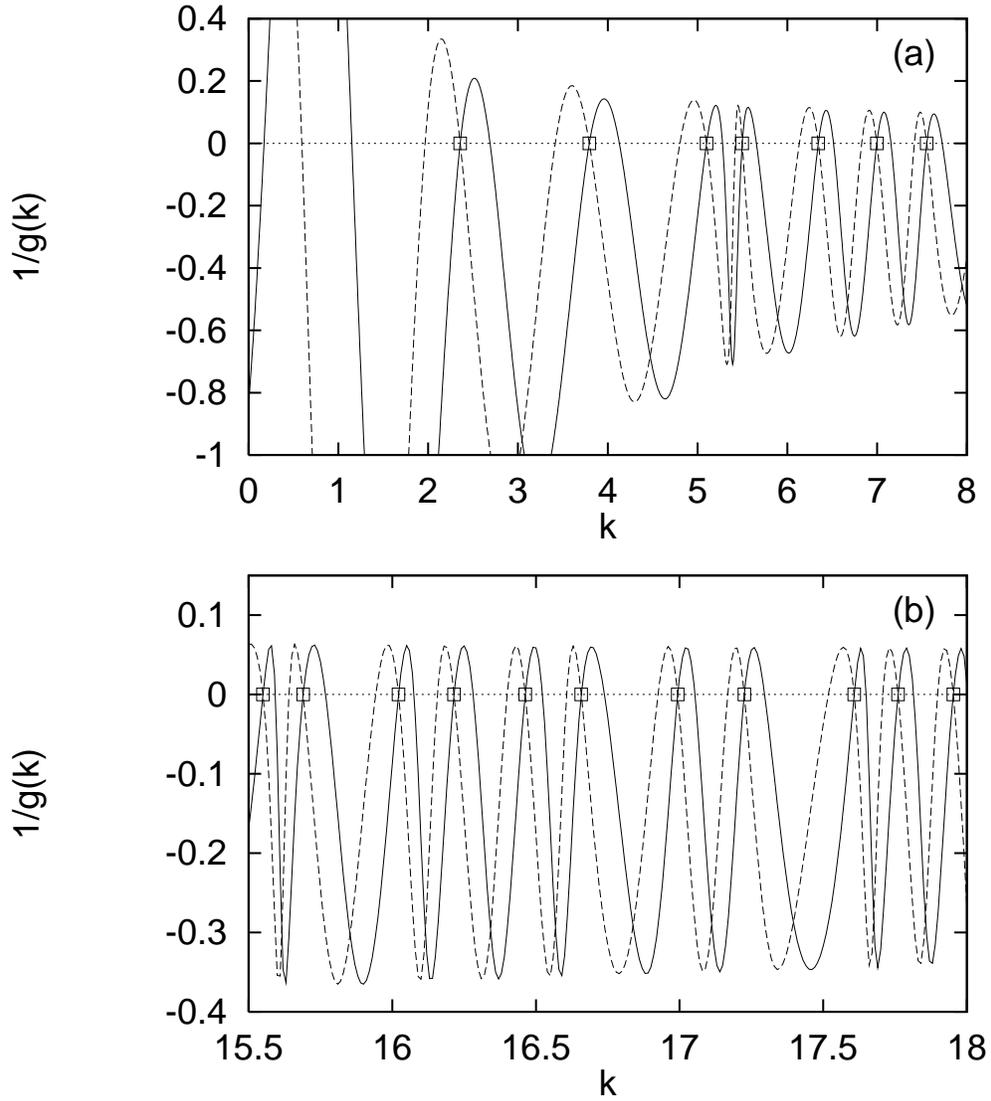}
\caption{
Real part (solid line) and imaginary part (dashed line) of the function
$1/g(k)$ for the circle billiard with radius $R=1$ obtained by Pad\'e 
approximant to the periodic orbit sum.
The zeros agree perfectly with the exact positions of the semiclassical
eigenvalues (from Eq.~\ref{EBK}) marked by the squares.
}
\label{fig5}
\end{figure}

It follows from Eq.~(\ref{A_circ:eq}) that the Pad\'e approximant to 
the periodic orbit sum should be calculated with an ordering parameter 
$n=m_r-1$, i.e.,
\begin{equation}
 A_n(k) = \sum_{m_r<n} {\cal A}_{\rm po}
   {\rm e}^{{\rm i}kL_{\rm po}} \; , \quad  n=1,2,\dots \; ,
\end{equation}
with the lengths $L_{\rm po}$ and amplitudes ${\cal A}_{\rm po}$ given 
by Eqs.\ (\ref{L_circ:eq}) and (\ref{A_circ:eq}), respectively.
We included all periodic orbits $(m_r,m_\varphi)$ with $m_r<100$ in the 
calculation of the function $g(k)$ which is the Pad\'e approximant to the 
sequence $A_n(k)$.
The real and imaginary parts of $1/g(k)$ are presented as solid and
dashed lines respectively in Fig.~\ref{fig5}.
The zeros of the function $1/g(k)$ agree perfectly to at least
seven significant digits with the exact positions of the
semiclassical eigenvalues obtained from Eq.~(\ref{EBK}) and marked
by the squares in Fig.~\ref{fig5}.

\section{Conclusion}
\label{concl:sec}
We have introduced two different methods to make semiclassical periodic orbit 
sums convergent.
The first method is based on the harmonic inversion of band-limited periodic 
orbit signals.
The characteristic feature of this method is the strict separation of the 
two steps, viz.\ firstly, the analytical filtering of the periodic orbit 
signal and, secondly, the numerical harmonic inversion of the band-limited 
signal by application of either linear predictor (LP), Pad\'e approximant 
(PA), or signal diagonalization (SD).
The separation of these two steps and the handling of small amounts of data
compared to other ``black box'' type signal processing techniques provides a
transparent approach to harmonic inversion as a semiclassical quantization
method thus opening the possibility to an easier and deeper understanding of
semiclassical quantization itself, and even yields numerically more accurate 
results than previous applications of filter-diagonalization (FD).
The harmonic inversion method can be applied to the periodic orbit 
quantization of systems with both chaotic and regular classical dynamics, 
when the periodic orbit signal is calculated with Gutzwiller's trace 
formula \cite{Gut67,Gut90} for isolated orbits and the Berry-Tabor 
formula \cite{Ber76} for orbits on rational tori, respectively.
More generally, any signal given as a sum of $\delta$ functions can be
filtered analytically and analyzed using the methods described in Sections
\ref{analyt_dec:sec} and \ref{hi:sec}.
For example, the technique can also be applied to the harmonic inversion 
of the density of states $\varrho(E)=\sum_n\delta(E-E_n)$ of quantum 
spectra to extract information about the underlying classical dynamics 
\cite{Mai99c,Mai97b,Gre00}.

The second method is the direct application of the Pad\'e approximant (PA)
to periodic orbit sums, and allows the resummation of the typically 
exponentially divergent terms of the semiclassical trace formulae.
The Pad\'e approximant can be applied when the total periodic orbit sum 
can be divided into partial sums with respect to an integer ordering 
parameter $n$, which can be related to, e.g., the cycle length of a 
symbolic code or the Maslov index of the orbit.

The two methods for periodic orbit quantization have been demonstrated for
systems with completely different classical dynamics, viz.\ the classically 
chaotic three-disk scattering problem and the integrable circle billiard.
A detailed comparison of various semiclassical quantization methods for the 
three-disk repeller reveals that quantization by harmonic inversion and
the Pad\'e approximant to periodic orbit sums can even be applied in energy 
regions where the cycle expansion of the Gutzwiller-Voros zeta function 
does not converge.

Evidently, the scope of applications of the methods presented in this paper
is wide, and runs from simple models to challenging atomic, molecular, and 
mesoscopic systems with typically mixed regular-chaotic classical dynamics.
Thus, even thirty years after the renaissance of semiclassics by Martin
Gutzwiller, new and unexpected insights may be gained into the no man's land
between classical and quantum physics.

\acknowledgements
Fruitful discussions with D\v z.\ Belki\'c, P.\ A.\ Dando, and H.\ S.\ Taylor
are gratefully acknowledged.
We thank A.\ Wirzba for supplying numerical data on the three-disk system.
This work was supported in part by the Deutsche Forschungsgemeinschaft
and Deutscher Akademischer Austauschdienst.

\newpage
\phantom{}

\newpage
\phantom{}

\newpage
\phantom{}

\newpage
\phantom{}

\end{document}